\documentclass[a4paper]{article}

\usepackage[english]{babel}
\usepackage[utf8]{inputenc}
\usepackage{amsmath}
\usepackage{graphicx}
\usepackage[colorinlistoftodos]{todonotes}
\usepackage[margin=0.75in]{geometry}
\usepackage{hyperref}

\newcommand{\beginsupplement}{%
        \setcounter{table}{0}
        \renewcommand{\thetable}{S\arabic{table}}%
        \setcounter{figure}{0}
        \renewcommand{\thefigure}{S\arabic{figure}}%
     }

\title{A glass half full interpretation of the replicability of psychological science}

\author{Jeffrey T. Leek, Prasad Patil, and Roger D. Peng}

\date{\today}

\begin{document}
\maketitle

\begin{abstract}
A recent study of the replicability of key psychological findings is a major contribution toward understanding the human side of the scientific process. Despite the careful and nuanced analysis reported in the paper, mass and social media adhered to the simple narrative that only 36\% of the studies replicated their original results. Here we show that 77\% of the replication effect sizes reported were within a prediction interval based on the original effect size. In this light, the results of \textit{Reproducibility Project: Psychology} can be viewed as a positive result for the scientific process. 
\end{abstract}

It is natural to hope that when two scientific experiments are conducted in the same way, they lead to identical conclusions. This is the intuition behind the recent tour-de-force replication of 100 psychological studies by the Open Science Collaboration, \textit{Reproducibility Project: Psychology}  \cite{open2015estimating}. At incredible expense and with painstaking effort, the researchers attempted to reproduce the exact conditions for each experiment, collect the data, and analyze them identically to the original study.

\begin{figure}[htp]
\begin{center}
\includegraphics[width=4.5in]{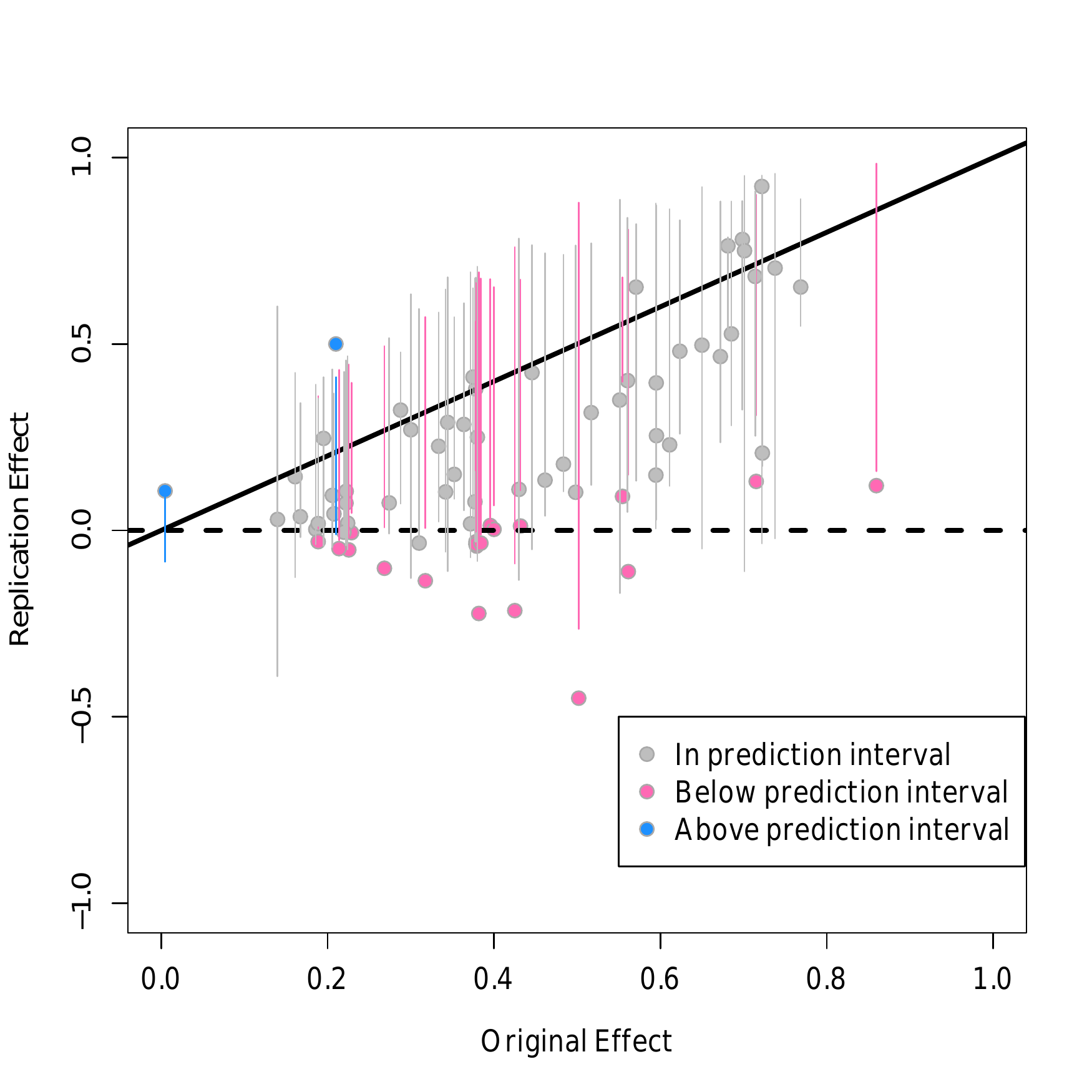}
\end{center}
\caption{\textbf{95\% prediction intervals suggest most replication effects fall in the expected range} A plot of original effects on the correlation scale (x-axis) and replication effects (y-axis). Each vertical line is the 95\% prediction interval based on the original effect size. Replication effects could either be below (pink), inside (grey), or above (blue) the prediction interval.}
\end{figure}

The original analysis considered both subjective and quantitative measures of whether the results of the original study were replicated in each case. They compared average effect sizes, compared effect sizes to confidence intervals, and measured subjective and qualitative assessments of replication. Despite the measured tone of the manuscript, the resulting mass and social media coverage of the paper fixated on a statement that only 36\% of the studies replicated the original result \cite{mediacoverage}.

This figure refers to the number of studies that reported a statistically significant ($P < 0.05$) result in both the original and replication studies. The relatively low number of results that were statistically significant in both studies was the focus of extreme headlines like ``Over half of psychology studies fail reproducibility test." \cite{baker2015half} and played into the prevailing narrative that science is in crisis \cite{gelman2014statistical}.

Unfortunately, the most widely publicized report from this paper is based on a misinterpretation of reproducibility and replicability.  Reproducibility is defined informally as the ability to recompute data analytic results conditional on an observed data set and knowledge of the statistical pipeline used to calculate them \cite{peng2011reproducible,peng2006reproducible}. The expectation for a study to be reproducible is that the exact same numbers will be produced from the same code and data every time.  Replicability of a study is the chance that a new experiment targeting the same scientific question will produce a consistent result \cite{asendorpf2013recommendations,ioannidis2005contradicted}. When a study is replicated, it is not expected that the same numbers will result for a host of reasons including both natural variability and changes in the sample population, methods, or analysis techniques \cite{leek2015statistics}. 

We therefore do not expect to get the same answer even if a perfect replication is performed.  Defining replication as consecutive results with $P < 0.05$ squares with the intuitive idea that replication studies should arrive at similar conclusions. So it makes sense that despite the many reported metrics in the original paper, the media has chosen to focus on this number. However, this definition is flawed since there is variation in both the original study and in the replication study. Even if you performed 100 perfect studies and 100 perfect replications of those studies, you would expect the number of times both P-values to be less than 0.05 to vary.

We conducted a small simulation based on the effect sizes presented in the original article. In the original study, the authors applied transformations to 73 of the 100 studies whose effects were reported via test statistics other than the correlation coefficient (e.g. t-statistics, F-statistics). We simulated 100 perfect replications of these 73 studies based on one degree of freedom tests. Each of these 100 simulations represents a perfect version of the Reproducibility Project with no errors. In each case, we calculated the fraction of P-values less than 0.05. The percentage of P-values less than 0.05 ranged from 66\% to 89\% with a high degree of variability (Figure S1).

Sampling variation alone may contribute to ``un-replicated" results if you define replication by a P-value cutoff. We instead consider a more direct approach by asking the question:``What effect would we expect to see in the replication study once we see the original effect?" This expectation depends on many variables about how the experiments are performed \cite{goodman1992comment}. Here we assume the replication experiment is indeed a true replication - a not unreasonable assumption in light of the effort gone to replicate these experiments accurately.

Assuming the replication is true and using the derived correlations from the original manuscript, we applied Fisher's z-transformation \cite{fisher1915frequency} to calculate a pointwise 95\% prediction interval for the replication effect size given the original effect. The 95\% prediction interval is $r_{orig} \pm \sqrt{\frac{1}{n_{orig}-3} + \frac{1}{n_{rep}-3}} z_{0.975}$, where $r_{orig}$ is the correlation estimate in the original study, $n_{orig}, n_{rep}$ are the sample sizes in the original and replication studies, and $z_{0.975}$ is the 97.5\% quantile of the normal distribution (Methods). A key difference between a confidence interval and the prediction interval is that the prediction interval accounts for variation in both the original study and in the replication study.

We observe that for the 92 studies where a replication correlation effect size could be calculated, 70 (or 75\%) were covered by the prediction interval based on the original correlation effect size (Figure 1). In two cases, the replication effect was actually larger than the upper bound of the prediction interval. Considering the asymmetric nature of the comparison, one might consider these effects as having ``replicated". We then estimate that 72/92 (or 77\%) of replication effects are in or above the prediction interval based on the original effect. Some of the effects that changed signs upon replication still fell within the prediction intervals calculated based on the original effects. This in unsurprising in light of the relatively modest sample sizes and effects in both the original and replication studies (Figure S2).

We also considered the 73 studies the author's reported to be based on one degree of freedom tests. In 51 of these 73 studies (70\%), the replication effect was within the prediction interval. The same two cases where the replication effect exceeded the prediction interval were in this set leaving us with an estimate of 53/73 (73\%) of these studies had replication effects consistent with the original effects. 

Based on the theory of the prediction interval we expect about 2.5\% of the replication effects to be above and 2.5\% of the replication effects to be below the prediction interval bounds. Since about 23\% were below the bounds, this suggests that not all effects replicate or that there were important sources of heterogeneity between the studies that were not accounted for.  The key message is that replication data---even for studies that should replicate---is subject to natural sampling variation in addition to a host of other confounding factors. 

We need a new definition for replication that acknowledges variation in both the original study and in the replication study. Specifically, a study replicates if the data collected from the replication are drawn from the same distribution as the data from the original experiment. To definitely evaluate replication we will need multiple independent replications of the same study. This view is consistent with the long-standing idea that a claim will only be settled by a scientific process rather than a single definitive scientific paper. 

The \textit{Reproducibility Project: Psychology} study highlights the fact that effects may be exaggerated and that replicating a study perfectly is challenging. We were caught off guard by the immediate and strong sentiment that psychology and other sciences may be in crisis~\cite{gelman2014statistical}. Our first reaction to Figure 3 from the original manuscript was pleasant surprise. The fact that many effects fall within the predicted ranges despite the long interval between original and replication study, the complicated nature of some of the experiments, and the differences in populations and investigators performing the studies is a reason for optimism about the scientific process. It is also in line with estimates we have previously made about the rate of false discoveries in the medical literature~\cite{jager2014estimate}. While there is a work to be done, the glass may not be quite as empty as the prevailing narrative would suggest.

\bibliographystyle{plain}
\bibliography{references}

\beginsupplement

\newpage

\section*{Methods}

\subsection*{Calculating a 95\% Prediction Interval}

We calculate a prediction interval based on the original $r_{orig}$ and replication $r_{rep}$ correlation estimates. Under normality and independence assumptions, the Fisher z-transformation provides the relationship: 

\begin{eqnarray*}
z^f = \frac{1}{2} \log \left(\frac{1+\hat{r}}{1-\hat{r}}\right) \sim N\left(\frac{1}{2}\frac{1+\rho}{1-\rho},\frac{1}{N-3}\right)
\end{eqnarray*}

Assume that $\hat{r}_{orig}$ and $\hat{r}_{rep}$ are the estimates from the original and replication studies and assume they have a common value. Then, we make the conservative assumption that the  original and replication experiments are independent, we can calculate:

\begin{eqnarray*}
\hat{z}^f_{orig} - \hat{z}^f_{rep} \sim N\left(0,\frac{1}{n_{orig}-3} + \frac{1}{n_{rep}-3}\right)\\
\end{eqnarray*}

then letting $se_{total} = \sqrt{\frac{1}{n_{orig}-3} + \frac{1}{n_{rep}-3}}$ we know that:

\begin{eqnarray*}
\frac{1}{se_{total}}(\hat{z}^f_{orig} - \hat{z}^f_{rep}) \sim N(0,1)
\end{eqnarray*}

so we have that:

\begin{eqnarray*}
Pr( z_{\alpha/2} < \frac{1}{se_{total}}(\hat{z}^f_{orig} - \hat{z}^f_{rep}) < z_{1-\alpha/2}) = \alpha\\
Pr( \hat{z}^f_{orig} - se_{total} z_{\alpha/2}  >  \hat{z}_{rep} > \hat{z}^f_{orig} - \sqrt{c} z_{1-\alpha/2} ) = \alpha
\end{eqnarray*}

So a $(1-\alpha)$\% prediction interval for $z^f_{rep}$ is $\hat{z}^f_{orig} \pm se_{total} z_{1-\alpha/2}$. We can then apply the inverse of the Fisher z-transform to obtain bounds for $r_{rep}$ on the appropriate scale.

\subsection*{P-value Simulation}

To simulate, we took all reported correlation coefficients for original studies. As described above, $arctan(\hat{r}) \sim N\left(\frac{1}{2}\frac{1+\rho}{1-\rho},\frac{1}{N-3}\right)$. We set $\rho = \hat{r}_{orig}$, and simulated 100 times, for each study, from the distribution $N\left(\frac{1}{2}\frac{1+\hat{r}_{orig}}{1-\hat{r}_{orig}},\frac{1}{N_{rep}-3}\right)$, where $N_{rep}$ was the sample size of each replication experiment. If $N_{rep}$ was unavailable, we used $N_{orig}$, and if both were unavailable, we used the median of the original sample sizes. If $\hat{r}_{orig}$ was unavailable, we similarly used the median of the correlations coefficients for the original studies.

Once we had 100 realizations from the distribution of the replicate correlation coefficients, we back-calculated them into F-statistics. We used the formula from the supplement of the original paper: $r = \sqrt{\dfrac{F\frac{df_1}{df_2}}{F\frac{df_1}{df_2} + 1}}\sqrt{\dfrac{1}{df_1}}$. From these 100 F-statistics, we were able to calculate 100 P-values and count up how many were $<$ 0.05

We made two assumptions/simplifications in the course of running this simulation, as it is merely for illustrative purposes. (1) We assumed that the correlation coefficient reported for the original study represented the true, population correlation coefficient (2) we converted all simulated correlation coefficients to $F(1, df_2)$ statistics, where $df_2$ were the degrees of freedom from the size of the replication study. Since 70\% of the original studies conducted the same analysis, we felt that this was a reasonable simplification for comparative purposes.

\subsection*{Code}

Code and data to reproduce this analysis is available from:

\begin{itemize}
\item \url{https://github.com/jtleek/replication_paper} 
\item \url{http://jtleek.com/replication_paper/code/replication_analysis.html}.  
\end{itemize}

\newpage

\section*{Supplementary Figures}

\begin{figure}[h]
\begin{center}
\includegraphics[width=6in]{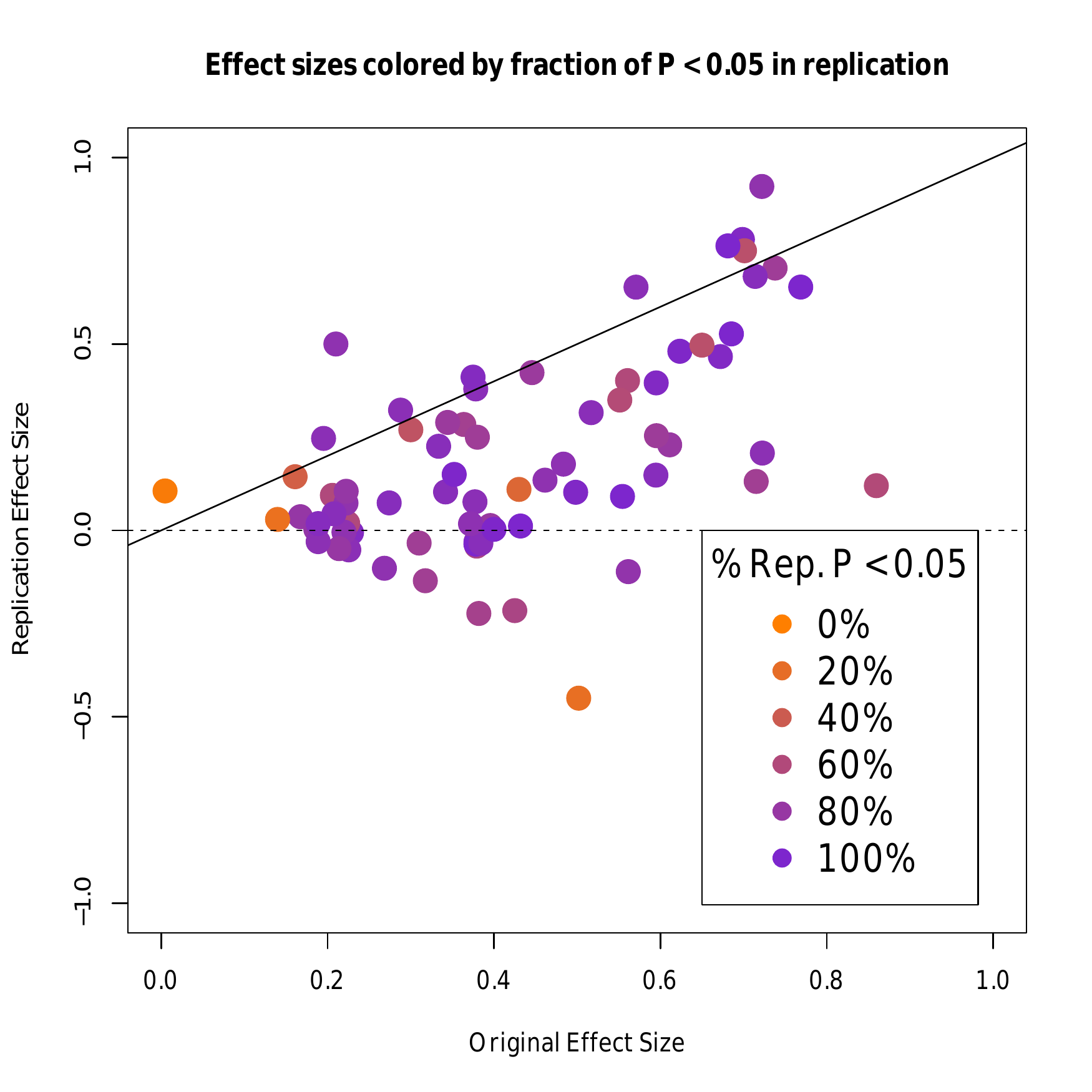}
\end{center}
\caption{\textbf{Empirical probability of replicating by effect size}. We simulated 100 effects from a distribution that assumes the original study effect is true. These were converted to test statistics, for which P-values were calculated. We then colored each point from Figure 3 in the original paper by how many times the calculated P-value was $<$ 0.05 out of the 100 simulations. This corresponds to the empirical probability of each study ``replicating'' by twice showing a statistically significant P-value}
\end{figure}

\begin{figure}[h]
\begin{center}
\includegraphics[width=6in]{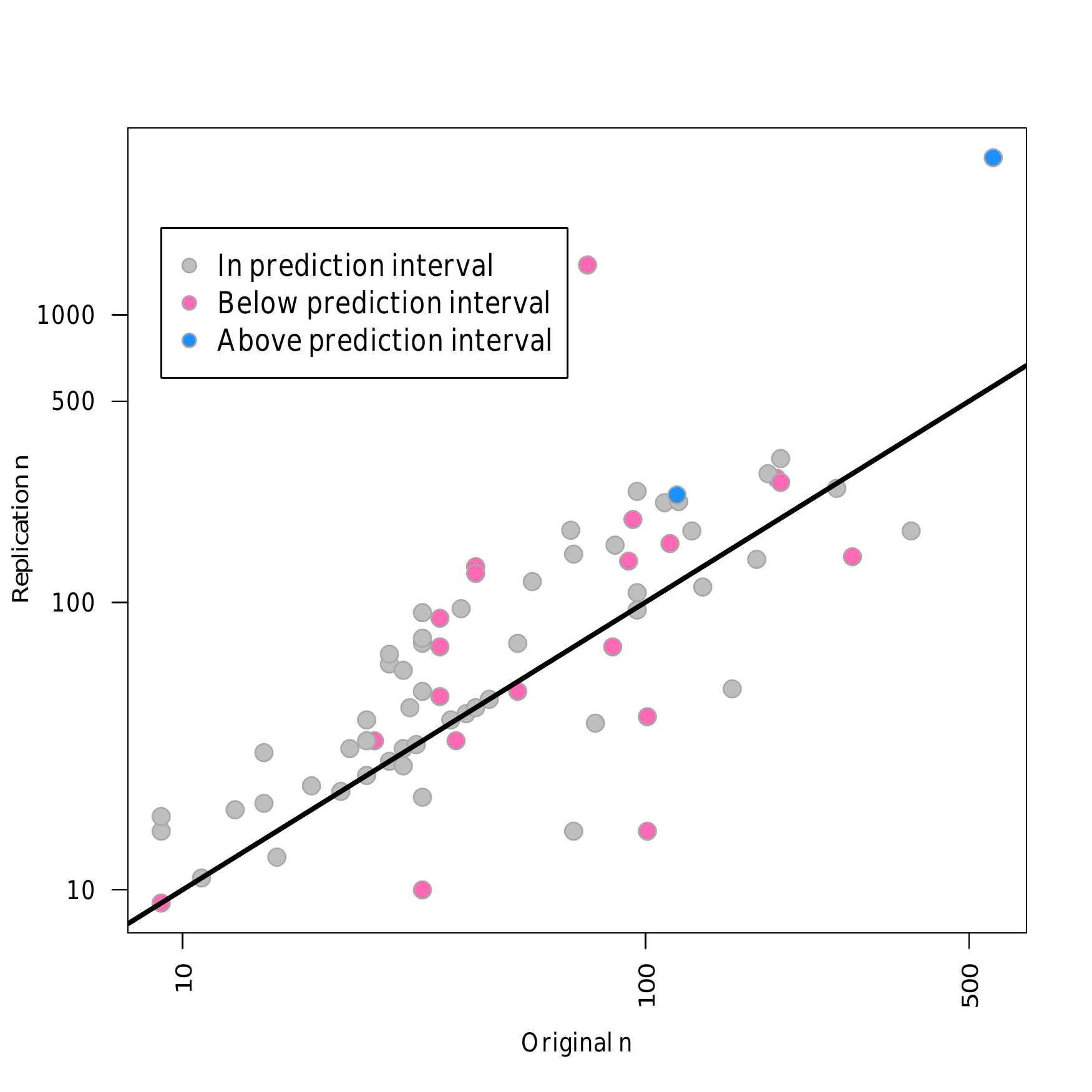}
\end{center}
\caption{\textbf{Sample sizes of studies in the Reproducibility Project colored by whether they fell in the 95\% prediction interval}. A plot of the original versus replication sample size colored by whether the resulting replication effect was inside (grey), above (blue) or below (pink) the 95\% prediction interval based on the original effect size. }
\end{figure}

\end{document}